\documentclass{elsart}
\usepackage{amssymb}
\usepackage{amsmath}
\usepackage{bm}
\usepackage{epsfig}
\usepackage[dvips]{color}
\expandafter\ifx\csname package@font\endcsname\relax\else
 \expandafter\expandafter
 \expandafter\usepackage
 \expandafter\expandafter
 \expandafter{\csname package@font\endcsname}%
\fi
\begin{document}
\begin{frontmatter}
\title{Gravimagnetic effect of the barycentric motion of the Sun and determination of the post-Newtonian parameter $\gamma$ in the Cassini experiment.}
\author[sk]{S.M. Kopeikin\corauthref{cor1}}\ead{kopeikins@missouri.edu}, \author[ap]{A.G. Polnarev}, \author[gos]{G. Sch\"afer}\and\author[iv]{I.Yu. Vlasov}
\address[sk]{Deptartment of Physics \& Astronomy, University of
Missouri-Columbia, Columbia, MO 65211, USA}
\address[ap]{Queen Mary University of London, London E14NS, UK}
\address[gos]{Institute of Theoretical Physics,  Friedrich-Schiller University, Jena D-07743, Germany}
\address[iv]{University of Guelph, Guelph N1G 2W1, Canada}
\corauth[cor1]{Corresponding author}
\begin{abstract}
The most precise test of the post-Newtonian $\gamma$ parameter in the solar
system has been achieved in measurement of the frequency shift of radio
waves to and from the Cassini spacecraft as they passed near the Sun.
The test relies upon the JPL model of radiowave propagation that includes, but does not explicitly
parametrize, the impact of the non-stationary component of the gravitational field
of the Sun, generated by its barycentric orbital motion, on the Shapiro delay. This non-stationary  gravitational
field of the Sun is associated with the Lorentz transformation of the metric tensor
and the affine connection from the heliocentric to the barycentric frame of
the solar system and can be treated as gravimagnetic field.
 The gravimagnetic field perturbs the propagation of a radio wave and
contributes to its frequency shift at the level up to $4\times 10^{-13}$ that may affect the precise measurement of the parameter $\gamma$ in the Cassini experiment to about one part in
10,000. Our analysis suggests that the translational gravimagnetic field of the Sun can be extracted from the Cassini data, and its effect is separable from the space curvature characterized by the parameter $\gamma$.
\end{abstract}
\maketitle\end{frontmatter}
\section{Introduction}
According to general relativity and other alternative theories of gravity, photons are deflected and delayed by the curvature of space-time produced by any mass distribution. In the gravitational field of a static spherical mass the bending and delay are proportional to $\gamma + 1$, where $\gamma$ parametrizes the curvature of space and measures the degree of deviation of a gravity theory from general relativity \cite{will}. The most precise measurement of $\gamma$ has been recently achieved in the Cassini experiment \cite{b0} by fitting the Doppler data to the general-relativistic model used in the JPL Orbit Determination Program \cite{moy}. The ODP calculates the time delay in the heliocentric frame and transforms it to the barycentric frame of the solar system by using spatial translations of participating bodies from the heliocentric to the barycentric frame. 

We demonstrate in this {\it Letter} that the spatial translations used in the ODP are equivalent to performing a Lorentz transformation of the solar gravitational field from the heliocentric (H) to the barycentric (B) frame. In the heliocentric frame photon moves in a static gravitational field of the Sun that is fully defined by the metric components $g^H_{00}$ and $g^H_{ij}$, while $g^H_{0i}=0$. The Lorentz transformation of the gravitational field to the barycentric frame brings about the gravimagnetic, $g^B_{0i}\not=0$, component of the metric tensor \cite{ki}. Moreover, it makes {\it all} components of the barycentric metric time-dependent, so that its partial time derivatives do not vanish. This non-stationarity of the solar gravity in the barycentric frame is associated with the gravimagnetic field caused by the translational current of Sun's matter \cite{cw} \footnote{Notice that GP-B experiment \cite{gpb} measures the gravimagnetic field caused by Earth's rotation mass-current that can not be generated by the Lorentz transformation.}.

This {\it Letter} shows that the barycentric orbital velocity of the Sun gives rise to gravimagnetic shift of frequency, $z_{\rm GM}\sim\alpha_\odot v_\odot/c$, where $\alpha_\odot$ is the solar gravitational deflection of light, $v_\odot\simeq 15$ m/s is the barycentric velocity of the Sun during the time of the Cassini experiment 2002, and $c$ is the speed of light. For the light grazing the solar limb $\alpha_\odot=8.5\times 10^{-6}$ rad, and $z_{\rm GM}\le 4.3\times 10^{-13}$. Therefore, detection of the gravimagnetic frequency shift $z_{\rm GM}$ is to be {\it possible} in ranging experiments like the Cassini having precision $10^{-14}$, or better. During the time radio signal travels from the spacecraft to the Earth the motion of the Sun can be approximated as uniform along a straight line as the solar acceleration, $a_\odot$, produces relativistic effect that is negligible for the Cassini but may be detected in future experiments \cite{lam}. This acceleration-induced frequency shift $z_{\rm acc.}\sim\alpha_\odot a_\odot R/c^2\simeq\alpha_\odot(R/R_\odot)(v_\odot/c)^2\ll 10^{-14}$, where $R$ is the distance spacecraft-observer, $R_\odot$ is the solar radius, and we have used $a_\odot\simeq v^2_\odot/R_\odot$ as the radius of the solar orbit around the barycentre is about $R_\odot$.

In what follows, the Greek indices $\alpha, \beta,...$ run from 0 to 3,
the Roman indices $i,j,...$ run from 1 to 3, repeated indices
mean Einstein's summation, and bold letters ${\bm
A}=(A^i), {\bm B}=(B^i),$ etc. denote spatial
vectors. A dot between two spatial vectors,
${\bm A}\cdot{\bm B}$, means the Euclidean scalar
product, and the cross between them, ${\bm
A}\times{\bm B}$, means the Euclidean vector product. The absolute value of a vector ${\bm A}$ is denoted $A=|{\bm A}|$.

\section{Gravitational Shift of Frequency in the Heliocentric Frame}

The heliocentric frame is denoted by $X^\alpha=(cT, {\bm X})$, where $T$ is time and ${\bm X}$ is the space coordinates with the Sun located at the origin. Spacecraft emits a radio signal at time $T_0$ at point ${\bm X}_0$, and observer (tracking station) receives this signal at time $T_1$ at point ${\bm X}_1$. The radio signal propagates along the direction defined by a unit vector ${\bm K}={\bm R}/R$, where ${\bm R}={\bm X}_1-{\bm X}_0$, and $R=|{\bm X}_1-{\bm X}_0|$. The total time of propagation of the radio signal from the spacecraft to the observer (with plasma dispersion eliminated) is
\begin{equation}
\label{-1}
T_1-T_0=\frac{R}{c}+\Delta T\;,
\end{equation}
where $\Delta T$ is the Shapiro gravitational time delay \cite{shap}  
\begin{equation}
\label{0}
\Delta T=(1+\gamma)\frac{GM_\odot}{c^3}\ln\left(\frac{R_0+R_1+R}{R_0+R_1-R}\right)\;,
\end{equation}   
$G$ is the universal gravitational constant, $M_\odot$ is mass of the Sun, $R_0=|{\bm X}_0|$, $R_1=|{\bm X}_1|$. Equation (\ref{0}) is valid in the post-Newtonian approximation and includes all relativistic terms of the order of $1/c^2$ in addition to the first term in the right side of equation (\ref{-1}). In the precise experiments like the Cassini some terms of the higher order of $1/c^4$ must be included to equation (\ref{0}). They were derived, for example, by \cite{rm2} and \cite{brum}. The most important of these higher-order terms are included to the ODP but are omitted here because we focus on the discussion of the velocity-dependent, gravimagnetic effects while the terms of the higher order are static. 

The ODP computes Doppler shift by differencing numerically the time delay equations (\ref{-1}), (\ref{0}) at the beginning, $T_0$, and the end, $T_1$, of the integration time. This procedure needs better theoretical understanding and is described here analytically in the approximation of a small impact parameter $D$ of the radio link that is sufficient for our subsequent discussion. In this approximation equation (\ref{0}) is simplified
\begin{equation}
\label{0+}
\Delta T=(1+\gamma)\frac{GM_\odot}{c^3}\ln\left(\frac{4L_0L_1}{D^2}\right)\;,
\end{equation} 
where $L_1\equiv({\bm K}\cdot{\bm X}_1)=\sqrt{R^2_1-D^2}$, $L_0\equiv -({\bm K}\cdot{\bm X}_0)=\sqrt{R^2_0-D^2}$, ${\bm D}\equiv{\bm K}\times({\bm X}_1\times{\bm K})={\bm K}\times({\bm X}_0\times{\bm K})$, $D=|{\bm D}|$, and we notice that $L_1+L_0=R$.  
 In the heliocentric frame, only the relative motion of the spacecraft and the Earth produce a spatial change in the impact parameter leading to a temporal change in the delay $\Delta T$ which re-calculates to a change in observed frequency $\nu_1$ of the radio signal emitted from the spacecraft with frequency $\nu_0$. The frequency change is $z=\nu_1/\nu_0-1=z_0+z_1+z_{\rm gr}$, where $z_0$ and $z_1$ are corrections accounting for transformation from time $T$ to the proper time $\tau$ at the spacecraft and observer positions respectively, and $z_{\rm gr}=d(\Delta T)/dT_1$ is the delay due to the variation in the strength of the gravitational field along the radio signal path due to the motion of observer, satellite, and/or the gravitating mass \cite{ks}. Only $z_{\rm gr}$ is important for analysis of the gravimagnetic effect of the barycentric motion of the Sun in precise determination of $\gamma$. Its calculation yields
\begin{equation}
\label{1}
z_{\rm gr}=-\frac{L_0}{cR}({\bm V}_1\cdot{\bm\alpha}_H)-\frac{L_1}{cR}({\bm V}_0\cdot{\bm\alpha}_H)+\delta z^H_{\rm gr}\;.
\end{equation} 
Here ${\bm V}_1=d{\bm X}_1/dT_1$ and ${\bm V}_0=d{\bm X}_0/dT_0$ are the heliocentric velocities of observer and emitter with respect to the Sun, residual terms 
\begin{equation}
\label{1az}
\delta z^H_{\rm gr}=\frac{\alpha_H}{2c}\left[\frac{D}{L_1}({\bm K}\cdot{\bm V}_1)-\frac{D}{L_0}({\bm K}\cdot{\bm V}_0)\right]+\frac{L_1}{c^2R}{\bm K}\cdot\left({\bm V}_1-{\bm V}_0\right)({\bm V}_0\cdot{\bm\alpha}_H)\;,
\end{equation}
where $\alpha_H=|{\bm\alpha_H}|$,
\begin{equation}
\label{2}
{\bm\alpha}_H=\alpha_\odot\,\frac{1+\gamma}{2}\frac{R_\odot}{D}\hat{\bm D}
\end{equation}
is a displacement vector describing magnitude and direction of the gravitational bending of the radio wave's trajectory, $\hat{\bm D}={\bm D}/D$. The asymmetry between subscripts 1 and 0 in the last term of equation (\ref{1az}) is because of the derivative $dT_0/dT_1=1-c^{-1}{\bm K}\cdot\left({\bm V}_1-{\bm V}_0\right)$ that appears any time when one differentiates functions depending on ${\bm X}_0={\bm X}(T_0(T_1))$ as $d{\bm X}_0/dT_1={\bm V}_0dT_0/dT_1$.

The first two terms in the right side of equation (\ref{1}) were obtained by \cite{b1}. However, the analytic theory of the Cassini experiment should also incorporate the term $\delta z^H_{\rm gr}$ given in equation (\ref{1az}). Indeed, the r.m.s. value of data points in the Cassini measurement corresponded to a one-way range rate of $\delta V=2.2\times 10^{-6}$ m/s \cite{b0}. Therefore, all terms having amplitude $\delta V/c\simeq 10^{-14}\div 10^{-15}$ must be kept in the analytic approach. The ODP calculates $\delta z^H_{\rm gr}$ numerically \cite{moy}. The analytic treatment of the gravitational Doppler shift provides deeper theoretical insight to the ultra-precise Doppler measurements in space navigation that is used for prediction more subtle relativistic effects \cite{ks,mash,kni,kkp} which are feasible for detection in future missions \cite{lam}.

\section{Gravitational Shift of Frequency in the Barycentric Frame}
\subsection{The Lorentz Transformation}\label{ltran}
The barycentric frame is denoted by $x^\alpha=(ct,{\bm x})$ where $t$ is a coordinate time and ${\bm x}$ are space coordinates with the origin located at the barycentre of the solar system \cite{soff}. Let ${\bm x}_\odot(t)$ denote the spatial coordinates of the Sun at time $t$ and its orbital velocity ${\bm v}_\odot(t)=d{\bm x}_\odot/dt$. Radio signal is emitted at ${\bm x}_0$ at time $t_0$, and received by observer at ${\bm x}_1$ at time $t_1$. Because we neglect the solar acceleration, ${\bm a}_\odot$, its coordinate can be approximated by uniform and rectilinear motion
\begin{equation}
\label{c1}
{\bm x}_\odot(t)={\bm x}_\odot+{\bm v}_\odot(t-t_A)+O\left[{\bm a}_\odot(t-t_A)^2\right]\;,
\end{equation}
where ${\bm x}_\odot\equiv{\bm x}_\odot(t_A)$ and ${\bm v}_\odot\equiv{\bm v}_\odot(t_A)$ are respectively the barycentric coordinate and velocity of the Sun taken at a fiducial instant of time, $t_A$, which is in the range between $t_0$ and $t_1$.

In what follows we assume that relativistic transformation of the solar gravitational field from the heliocentric to the barycentric frame can be performed with the matrix of the Lorentz transformation $\Lambda^\alpha_\beta$ having a standard special-relativistic form \cite{mtw} 
\begin{eqnarray}
\label{mat1}
\Lambda^0_{\;0}=\Gamma\;,\qquad\qquad
\Lambda^0_{\;i}=\Lambda^i_{\;0}=-\Gamma\beta^i\;,\qquad\qquad
\Lambda^i_{\;j}=\delta^{ij}+\left(\Gamma-1\right)\frac{\beta^i\beta^j}{\beta^2}\;,
\end{eqnarray}
where $\Gamma\equiv\sqrt{1-\beta^2}$, and $\beta^i=v_\odot^i/c\ll 1$. This matrix transforms coordinates as follows
\begin{equation}
\label{c2}
X^\alpha=\Lambda^\alpha_\beta\,(x^\beta-x_\odot^\beta)\;,
\end{equation}
where $x_\odot^\beta=(ct_A,{\bm x}_\odot)$. The Lorentz transformation transfrorms the metric tensor and the affine connection, and make them time-dependent in the barycentric frame of reference. All time-dependent terms that are proportional to the barycentric velocity of the Sun, $v_\odot$, are called gravimagnetic \cite{cw} and are associated with the gravimagnetic effect in the Doppler shift. Precise equations describing the Lorentz transformation of the gravity field and their connection to the propagation of light are given in \cite{ki}. 

As we are interested only in the gravimagnetic shift of frequency of the first order, it suffices to approximate equation (\ref{c2}) by making use of the Taylor expansion of $\Lambda^\alpha_\beta$ with respect to ${\bm\beta}$ and to work with a few first terms. This reduces equation (\ref{c2}) to the post-Galilean transformation 
\begin{eqnarray}
\label{5}
{\bm X}&=&{\bm x}-{\bm x}_\odot-{\bm v}_\odot (t-t_A)+O(v_\odot^2/c^2)\;,\\
\label{6}
T&=&t+O(v_\odot/c^2)\;.
\end{eqnarray}
It is important to emphasize that the term proportional to $v_\odot$ in equation (\ref{5}) is, effectively, a linear order with respect to $v_\odot/c$, as it is obtained from the contraction of $\Lambda^i_{\;0}$ from equation (\ref{mat1}) with time coordinate $x^0=ct$. Therefore, the Lorentz invariance of the gravity field reveals itself already at the level of the post-Galilean transformation (\ref{5}) that can be tested experimentally with the Cassini data.

\subsection{Transformation of the Gravitational Shift of Frequency }   
Transformation of the frequency shift equation (\ref{1}) to the barycentric frame requires transformation of the heliocentric distances $R_0$, $R_1$, $R$, the impact parameter ${\bm D}$, and velocities ${\bm V}_0$, ${\bm V}_1$ to their barycentric counterparts. By making use of equations (\ref{5}), (\ref{6}) we obtain
\begin{eqnarray}
\label{n1}
{\bm X}_1&=&{\bm r}_1\equiv {\bm x}_1-{\bm x}_\odot(t_1)\;,\\
\label{n2}
{\bm X}_0&=&{\bm r}_0\equiv {\bm x}_0-{\bm x}_\odot(t_0)\;,
\\
\label{7a}
{\bm V}_1&=&{\bm v}_1-{\bm v}_\odot\;,\\\label{8a}
{\bm V}_0&=&{\bm v}_0-{\bm v}_\odot\;,
\end{eqnarray}
where ${\bm x}_\odot(t_0)$ and ${\bm x}_\odot(t_1)$ are defined by equation (\ref{c1}), ${\bm v}_1=[d{\bm x}_1(t)/dt]_{t=t_1}$ and ${\bm v}_0=[d{\bm x}_0(t)/dt]_{t=t_0}$ are the barycentric velocities of observer and emitter of the radio signal respectively. 

Equations (\ref{n1}), (\ref{n2}) are the spatial translations used in the ODP for calculating the barycentric time delay \cite{moy}. They allow us to derive transformations of the distance ${\bm R}$ and the unit vector ${\bm K}$. Specifically,
\begin{eqnarray} 
\label{n3}
{\bm R}&=&{\bm r}-\frac{{\bm v}_\odot}{c}\,r\;,\\
\label{n4}
{\bm K}&=&{\bm k}-\frac{{\bm v}^\perp_\odot}{c}\;,
\end{eqnarray}
where ${\bm r}={\bm x}_1-{\bm x}_0$, $r=|{\bm r}|$, the unit vector ${\bm k}\equiv({\bm x}_1-{\bm x}_0)/|{\bm x}_1-{\bm x}_0|$ defines direction of propagation of the radio signal in the barycentric frame, and ${\bm v}^\perp_\odot\equiv{\bm k}\times\left({\bm v}_\odot\times{\bm k}\right)$ is the Sun's velocity projected on the plane being orthogonal to ${\bm k}$.
Transformation of the heliocentric vector ${\bm D}$ is given by equation \cite{klio}
\begin{equation}
\label{n5}
{\bm D}={\bm d}+({\bm d}\cdot{\bm v}_\odot){\bm k}+({\bm k}\cdot{\bm r}_1){\bm v}^\perp_\odot\;,
\end{equation}
where ${\bm d}\equiv{\bm k}\times({\bm r}_1\times{\bm k})$ is the barycentric impact parameter vector of the light ray taken at instant $t_1$. 
Using equation (\ref{n5})
the deflection angle ${\bm\alpha}_H$ is transformed as follows \cite{klio}
\begin{eqnarray}
\label{n6}
{\bm\alpha}_H&=&{\bm\alpha}_B+({\bm\alpha}_B\cdot{\bm v}_\odot){\bm k}+\frac{({\bm k}\cdot{\bm r}_1)}{d}\left[\alpha_B{\bm v}^\perp_\odot -2({\bm d}\cdot{\bm v}_\odot){\bm\alpha}_B\right]\;,\\  
\label{n7}
{\bm\alpha}_B&=&\alpha_\odot\,\frac{1+\gamma}{2}\frac{R_\odot}{d}\hat{\bm d}\;,
\end{eqnarray}
where $\alpha_B=|{\bm\alpha_B}|$, and the unit vector $\hat{\bm d}={\bm d}/|{\bm d}|$.

Substitution of the heliocentric quantities (\ref{n1})--(\ref{n7}) to the frequency shift equation (\ref{1}) yields
\begin{equation}
\label{3}
z_{\rm gr}=-\underbrace{\quad\frac{l_0}{cr}({\bm v}_1\cdot{\bm\alpha}_B)\quad}_{\rm observer\;shift\;z_{\rm O}}-\underbrace{\quad\frac{l_1}{cr}({\bm v}_0\cdot{\bm\alpha}_B)\quad}_{\rm satellite\;shift\;z_{\rm S}}+\underbrace{\qquad\frac{1}{c}({\bm v}_\odot\cdot{\bm\alpha}_B)\qquad}_{\rm gravimagnetic\;shift\;z_{\rm GM}}+\delta z^B_{\rm gr}\;,
\end{equation}
where $l_0\equiv-({\bm k}\cdot{\bm r}_0)$, $l_1\equiv({\bm k}\cdot{\bm r}_1)$, and $\delta z^B_{\rm gr}$ denotes the residual terms with magnitude up to $10^{-16}$ \cite{ks}. In the Cassini experiment the distances $l_0\simeq 7.4$ AU, $l_1\simeq 1.0$ AU, and $r=8.43$ AU.
\subsection{Parametrization of the Shapiro Delay in the Barycentric Frame}\label{psd}

Linearized transformations (\ref{n1}) and (\ref{n2}) of the Shapiro time delay are used in the ODP \cite{moy} but they are not parametrized to track down the gravimagnetic effect of the solar barycentric velocity {\it explicitly}, which may be also important for other, ultra-precise solar system experiments \cite{lam}. The most simple way to introduce such a parametrization is to label velocity of the Lorentz transformation by replacing $\beta^i\rightarrow\epsilon\beta^i$, where $\epsilon$ is a new fitting parameter \cite{ki}. It is not equivalent to the PPN parameter $\alpha_1$ as $\alpha_1$ is associated with the ``preferred-frame" effects \cite{will} which are absent in general relativity, while $\epsilon$ marks genuine gravimagnetic effects induced by the Lorentz transformation of gravitational field \cite{kopfom}. The Cassini data are accurate enough to measure $\epsilon$, thus, providing a test of the existence of the gravimagnetic field caused by the barycentric motion of the Sun. Gaussian distribution of the resiuduals of the post-fit Cassini data \cite{b0} signals that $\epsilon$ is close to unity. However, it is tantalizing to obtain its precise statistical estimate that will provide a new test of general relativity at higher technological level.  

The suggested $\epsilon$-parameterization recasts equations (\ref{n1}), (\ref{n2}) to the following form 
\begin{eqnarray}
\label{c3}
{\bm X}_1&=&{\bm r}_1\equiv {\bm x}_1-{\bm x}_\odot(\epsilon t_1)\;,\\
\label{c4}
{\bm X}_0&=&{\bm r}_0\equiv {\bm x}_0-{\bm x}_\odot(\epsilon t_0)\;,
\end{eqnarray}
so that equation (\ref{3}) becomes
\begin{equation}
\label{3z}
z_{\rm gr}=-\frac{l_0}{cr}({\bm v}_1\cdot{\bm\alpha}_B)-\frac{l_1}{cr}({\bm v}_0\cdot{\bm\alpha}_B)+\frac{\epsilon}{c}({\bm v}_\odot\cdot{\bm\alpha}_B)+\delta z^B_{\rm gr}(\epsilon)\;,
\end{equation} 
making evident that $\epsilon$ traces that part of the Doppler shift which is solely due to the barycentric orbital motion of the Sun. 
The partial derivatives of the Shapiro delay with respect to parameters $\gamma$ and $\epsilon$ are
\begin{eqnarray}
\label{p1}
\left[\frac{\partial\Delta T}{\partial\gamma}\right]_{\epsilon=1}&=&\frac{GM_\odot}{c^3}\ln\left[\frac{r_0+r_1+r-({\bm r}\cdot{\bm v}_\odot)/c}{r_0+r_1-r+({\bm r}\cdot{\bm v}_\odot)/c}\right]\;,\\
\label{p2}
\left[\frac{\partial\Delta T}{\partial\epsilon}\right]_{\gamma=1}&=&\frac{4GM_\odot}{c^3}\frac{r}{(r_1+r_0)^2-r^2}\,{\bm v}_\odot\cdot\left({\bm k}_1t_1+{\bm k}_0t_0-\frac{{\bm k}}{c}(r_1+r_2)\right),
\end{eqnarray} 
where ${\bm k}_1={\bm r}_1/r_1$ and ${\bm k}_0={\bm r}_0/r_0$. Statistical meaning of partial derivatives (\ref{p1}) and (\ref{p2}) is that they are components of the covariance matrix of the fitting parameters. The fact, that they have different functional dependence, points out that parameters $\gamma$ and $\epsilon$ are uncorrelated and, hence, can be separated with confidence in the data fitting procedure.

\section{Discussion}

 The Cassini experiment was conducted in June 2002 when radio waves from/to the spacecraft had passed at the minimal distance $d=1.6$ $R_\odot$ from the Sun. The peak value of $[z_{\rm gr}]_{peak}\simeq\alpha_\odot (R_\odot/d)(v_\oplus/c)\simeq 6\times 10^{-10}$ is caused by the orbital velocity $v_\oplus$ of the Earth, while the measurement errors of $z_{\rm gr}$ were about five orders of magnitude smaller \cite{b0}. 

\begin{figure}
\centerline{\epsfig{figure=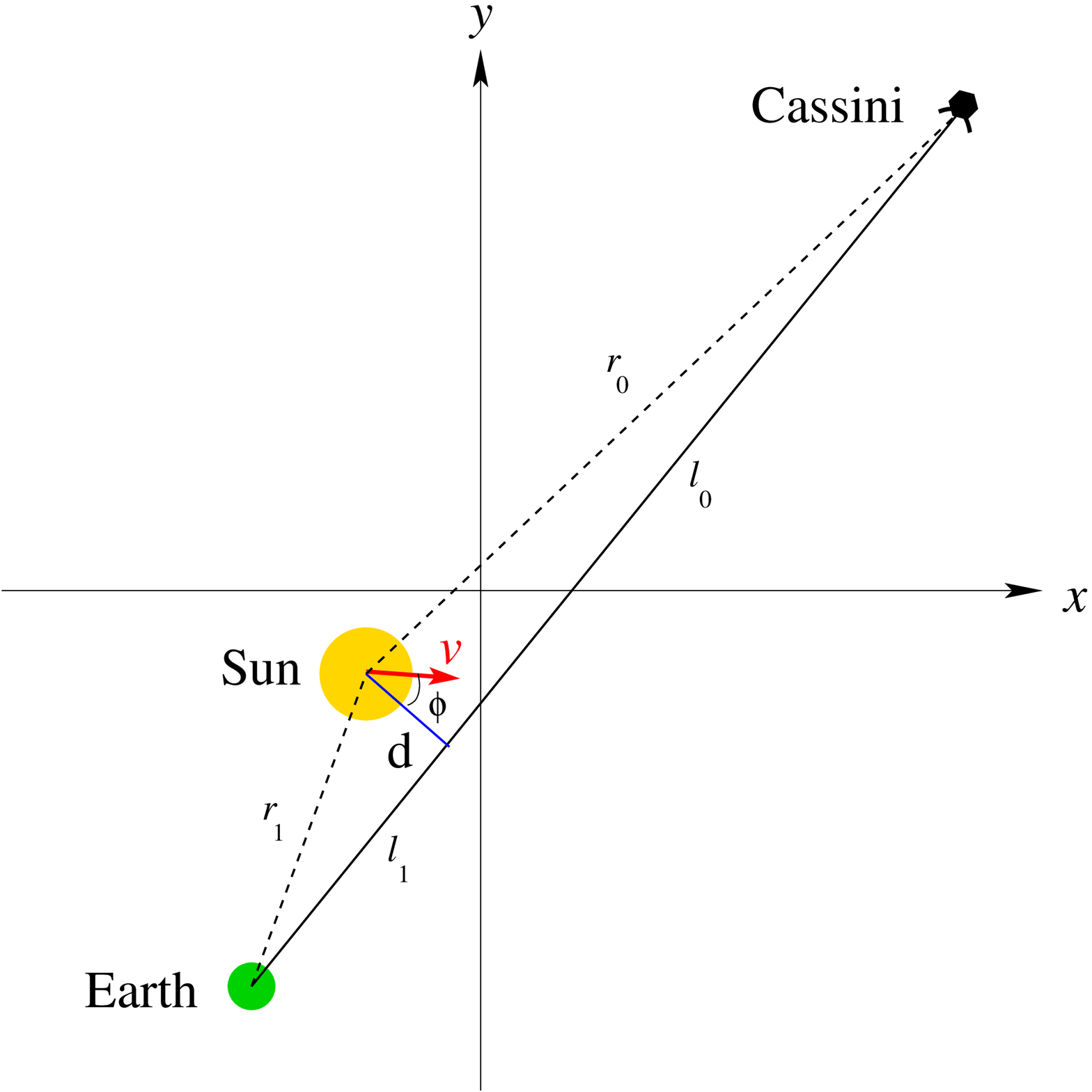}}
\vspace{1cm}
\caption{Relative configuration of the Cassini, the Sun and the Earth by the time of solar conjunction on June 21, 2002 in the barycentric reference frame with origin located at the centre of mass of the solar system. The barycentric velocity of the Sun is shown by (red) arrow that makes an angle $\phi\simeq 16^\circ$ with respect to the impact parameter of the radio-link connecting the Cassini spacecraft and the Doppler tracking station on the Earth. This angle changes to $\phi\simeq 164^\circ$ after the radio-link gets on the other side from the Sun.\label{fig2}}
\end{figure}

We have used JPL ephemerides to calculate the barycentric position and velocity of the Earth and the Sun at the time of the Cassini-solar conjunction. One of the consecutive configuration of these bodies is shown in Fig. \ref{fig2}. It varies rapidly as all three bodies are moving with respect to each other, thus, leading to the change of the impact parameter of the radio link.
The solar barycentric velocity is about 14.4 m/s, and the angle $\phi$ between the direction of ${\bm v}_\odot$ and the vector ${\bm d}$ is about $16^\circ$. This angle changes its value to $\phi\simeq 164^\circ$ after the satellite passes on the other side of Sun's limb, thus, changing the sign of the gravimagnetic effect.

These values were used for comparative estimation of the gravimagnetic term in equation (\ref{3}). The Doppler residuals in the experiment run by \cite{b0}, were obtained from 9-parameter fit on the calibrated data with the solar corona and the interplanetary medium dispersions removed with double-frequency observations. The r.m.s. value of the $N=1094$ data points (relative to an 8.4 GHz carrier) is $1.2\times 10^{-4}$ Hz, corresponding to a one-way range rate of $2.2\times 10^{-6}$ m s$^{-1}$ or for a relative frequency shift $\delta z=0.7\times 10^{-14}$. It means that the average error of a single measurement is $\sigma_z=\sqrt{2N}\delta z\simeq 3.2\times 10^{-13}$. The gravimagnetic frequency shift caused by the barycentric orbital motion of the Sun amounts to the peak value of $[z_{\rm GM}]_{peak}=(r/l_0)(v_\odot/v_\oplus)\cos\phi\,[z_{\rm gr}]_{peak}\simeq 3.3\times 10^{-13}$ that is comparable with the r.m.s. of a single measurement but a comprehensive number of measurements is required to extract $z_{\rm GM}$ out of noise. In order to estimate the number of the measurements being available for this goal from the existing data, we notice that the gravimagnetic term $z_{\rm GM}\sim(v_\odot/c)\alpha_B$ reaches the Doppler data r.m.s. value for $d\simeq 30$ $R_\odot$ that corresponds to 20 days of observations around the solar conjunction giving the number of data points $n\simeq 800$. This should be statistically sufficient to measure the gravimagnetic effect caused by the barycentric motion of the Sun. 

Equation (\ref{3}) suggests that the gravimagnetic term $z_{\rm GM}$, if it were ignored for whatever reason, would affect the measured value of the parameter $\gamma$. The bias in $\gamma$ is approximately given by the ratio $[z_{\rm GM}]_{peak}/[z_{\rm gr}]_{peak}\le 5\times 10^{-4}$ which should be further relaxed by a factor of $3\div 5$ for $z_{\rm GM}$ is extracted from a smaller subset of the entire data \footnote{\cite{kostel} have arrived to similar numerical estimate by analyzing solar-system experiments in the framework of the Standard Model Extensions. }. This value is larger than $\gamma-1=(+2.1\pm 2.3)\times 10^{-5}$ given by \cite{b0} pointing out that the gravimagnetic field due to the solar barycentric velocity $v_\odot$ can be measured though its extraction may represent difficulties (Bertotti 2006, private communication).

This {\it Letter} suggests to disentangle the measurement of the PPN parameter $\gamma$ in the Cassini experiment \cite{b0,a04} from the effect of the gravimagnetic field caused by the barycentric orbital motion of the Sun. The orbital motion must be properly parametrized in the delay equation (\ref{0}) by making use of the translations (\ref{c3}), (\ref{c4}) where $\epsilon$ is a fitting parameter. This paper calculates the gravimagnetic frequency shift in one-way radio link. In real experiment two -- up and down -- radio links are used. Analytic equations of the Doppler shift for the two radio links were derived in  \cite{ks} and they substantiate our conclusions. 

The translationally-invariant gravimagnetic field can be measured more accurately in space missions LATOR \cite{lator} and ASTROD \cite{astrod}. The space interferometer SIM \cite{sim} and ground-based Square Kilometer Array \cite{ska} can also achieve this goal \cite{kopmak}.     
     
\section*{Acknowledgments}
We thank J.~D. Anderson and B. Bertotti for valuable discussions, and P. Bender for instructive remarks.

\clearpage

\begin{thebibliography}{99}
\bibitem[1]{will} Will, C.~M., 1993, {\it Theory and experiment in gravitational physics} (Cambridge University Press: Cambridge) 
\bibitem[2]{b0} Bertotti, B., Iess, L. \& Tortora, P., 2003, {\it Nature}, {\bf 425}, 374
\bibitem[3]{moy} Moyer, T.~D., 2003, {\it Formulation for Observed and Computed Values of Deep Space Network Data Types for Navigation} (John Wiley \& Sons: Hoboken)
\bibitem[4]{ki} Kopeikin, S.~M., 2006, {\it Int. J. Mod. Phys.}, {\bf D15}, 305
\bibitem[5]{cw} Ciufolini, I. \& Wheeler,  J.A., 1995, {\it Gravitation and Inertia} (Princeton University Press: Princeton)
\bibitem[6]{gpb} Muhlfelder, B., 2005, {\it Gravity Probe B}, APS Meeting Abstracts, \#J4.002
\bibitem[7]{lam} Dittus, H., L\"ammerzahl, C. \& Turyshev, S.~G., 2006,  {\it Lasers, Clocks, and Drag-Free: Exploration of Relativistic Gravity in Space}, (Springer: Berlin) 
\bibitem[8]{shap} Shapiro, I.~I., 1964, {\it Phys. Rev. Lett.}, {\bf 13}, 789
\bibitem[9]{rm2}Richter, G.~W. \& Matzner, R.~A., \ 1983, {\it Phys. Rev. D}, {\bf 28}, 3007 
\bibitem[10]{brum} Brumberg, V.~A., 1991, {\it Essential Relativistic Celestial Mechanics} (Adam Hilger: Bristol)
\bibitem[11]{ks} Kopeikin, S.~M. \& Sch\"afer, G.,  1999, {\it Phys. Rev D}, {\bf 60}, 124002 
\bibitem[12]{b1} Bertotti, B. \& Giampieri, G., 1992,  {\it Class. Quant. Grav.}, {\bf 9}, 777
\bibitem[13]{mash} Kopeikin, S. \& Mashhoon, B.,\ 2002, {\it Phys. Rev. D}, {\bf 65}, 064025
\bibitem[14]{kni} Kopeikin, S.~M. \& Ni, 
W.-T.,\ 2006, in: {\it Lasers, Clocks, and Drag-Free: Exploration of Relativistic Gravity in Space}, Eds. H. Dittus, C. L\"ammerzahl \& S.~G. Turyshev (Springer: Berlin), pp. 209--216

\bibitem[15]{kkp} Kopeikin, S., Korobkov, P. \& Polnarev, A., 2006, {\it Class. Quantum Grav.}, {\bf 23}, 4299
\bibitem[16]{soff}Soffel, M., Klioner, S. A., Petit, G., et al.,  2003, {\it Astron. J.}, {\bf 126}, 2687
\bibitem[17]{mtw} Misner, C.~W., Thorne, K.~S. \& Wheeler, J.~A., \ 1973, {\it Gravitation} (New York: Freeman)
\bibitem[18]{klio} Klioner,  S.~A., 2003, {\it Astron. Astrophys.}, {\bf 404}, 783
\bibitem[19]{kopfom}Kopeikin, S.~M. \& Fomalont, E.~B.,\ 2006, {\it Found. of Physics}, {\bf 36}, 1244 
\bibitem[20]{kostel} Bailey, Q.~G. \& Kosteleck{\'y}, V.~A.,\ 2006, {\it Phys. Rev. D}, {\bf 74}, 045001 
\bibitem[21]{a04}Anderson, J.~D., Lau, E.~L. \& Giampieri, G., 2004, in: Proc. of the 22nd Texas Symp. on Rel. Astrophys., Stanford, C041213, 0305 
\bibitem[22]{lator}
Turyshev, S.~G., Shao, M., \& Nordtvedt, K.~L.,\ 2006, in: {\it Lasers, Clocks, and Drag-Free: Exploration of Relativistic Gravity in Space}, Eds. H. Dittus, C. L\"ammerzahl \& S.~G. Turyshev (Springer: Berlin), pp. 429-493
\bibitem[23]{astrod} Ni, W.-T., Bao, Y., Dittus, H. et al.,\ 2006, {\it Acta Astronautica}, {\bf 59}, 598
\bibitem[24]{sim} Shao, M., \ 2006, In: {\it Advances in Stellar Interferometry}, Eds. J.~D. Monnier, M.  Sch\"oller \& W.~C. Danchi. Proc. of the SPIE, vol. 6268, p. 61 
\bibitem[25]{ska} Beck, R.,\ 2005, {\it Astronomische 
Nachrichten}, {\bf 326}, 608 
\bibitem[26]{kopmak}Kopeikin, S.~M. \& Makarov, V.~V.,\ 2007, {\it Phys. Rev. D}, in press; e-print: astro-ph/0611358 
\end{thebibliography}
\end{document}